%% file: main.tex
\pdfoutput=1

\documentclass[11pt]{article}

\usepackage[final]{acl}

\usepackage{times}
\usepackage{latexsym}
\usepackage{array}
\usepackage{booktabs}
\usepackage{enumitem} 
\usepackage{listings}
\usepackage{xcolor}
\usepackage{amsmath}
\usepackage{footmisc}

\definecolor{promptbg}{RGB}{245,248,255}
\definecolor{promptframe}{RGB}{180,180,180}
\usepackage[T1]{fontenc}

\usepackage[utf8]{inputenc}

\usepackage{microtype}

\usepackage{inconsolata}

\usepackage{graphicx}

\title{SWE-MERA: A Dynamic Benchmark for Agenticly Evaluating Large Language Models on Software Engineering Tasks}

\author{
 \textbf{Pavel Adamenko\textsuperscript{1}},
 \textbf{Mikhail Ivanov\textsuperscript{2}},
 \textbf{Aidar Valeev\textsuperscript{1}},\\
 \textbf{Rodion Levichev\textsuperscript{1}},
 \textbf{Pavel Zadorozhny\textsuperscript{1}},
 \textbf{Ivan Lopatin\textsuperscript{1}},\\
\textbf{Dmitrii Babaev\textsuperscript{1}},
 \textbf{Alena Fenogenova\textsuperscript{1}},
 \textbf{Valentin Malykh\textsuperscript{3,2,4}}
\\
\\
 \textsuperscript{1}GigaCode,
 \textsuperscript{2}ITMO University,
  \textsuperscript{3}MWS AI,
  \textsuperscript{4}IITU university,
\\
 \small{
   \textbf{Correspondence:} \href{mailto:mera@a-ai.ru}{mera@a-ai.ru}
 }
}

\begin{document}
\maketitle
\begin{abstract}
The rapid advancement of Large Language Models (LLMs) in software engineering has revealed critical limitations in existing benchmarks, particularly the widely used SWE-bench dataset. Recent studies have uncovered severe data contamination issues, e.g., SWE-bench~\cite{jimenez2023swe} reports 32.67\% of successful patches involve direct solution leakage and 31.08\% pass due to inadequate test cases. We introduce \textbf{SWE-MERA}, a dynamic, continuously updated benchmark designed to address these fundamental challenges through an automated collection of real-world GitHub issues and rigorous quality validation. Our approach implements a reliable pipeline that ensures quality while minimizing contamination risks, resulting in approximately 10,000 potential tasks with 728 samples currently available. Evaluation using the Aider coding agent demonstrates strong discriminative power in state-of-the-art models. We report performance across a dozen recent LLMs evaluated on tasks collected between September 2024 and June 2025.
\end{abstract}

\section{Introduction}
The complexity of real-world software development processes goes beyond merely completing code. It encompasses coding agents and a range of text-to-code tasks. E.g., SWE-bench~\cite{jimenez2023swe} was created from a dataset comprising 2,294 GitHub issues and their corresponding pull requests (PRs). Each task in SWE-bench represents an authentic, real-world problem structured around: 1) The initial commit (code before changes),
2) The fixing commit (solution to the problem), 3) The issue description (what needed to be fixed).
A critical limitation of this benchmark is its \textit{static nature} — the tasks were collected only once and never updated. This leads to two major issues.
\textit{Data leakage}: As models are repeatedly tested on the same fixed dataset, they may inadvertently memorize solutions or overfit to outdated examples. \textit{Benchmark saturation}: Over time, the benchmark loses its effectiveness as state-of-the-art models achieve near-perfect scores, making it harder to distinguish meaningful progress.


\textbf{SWE-MERA} addresses these shortcomings (typical for many code benchmarks) by introducing \textit{dynamic updates} to the test cases. Regularly refreshing the dataset with new, unseen issues ensures:
1) \textit{real-world relevance} — tasks reflect the latest challenges in software development
2) \textit{fair evaluation} — models are tested on fresh problems, minimizing the risk of data leakage
3) \textit{continuous improvement} — the benchmark evolves in tandem with advancements in AI and software engineering practices.

The contributions of the paper are as follows:

\begin{enumerate}[nosep]
    \item The seven-stage pipeline effectively ensures quality and minimizes contamination risks, able to collect approximately 10,000 potential tasks, with 728 samples currently available.
    \item   An automated scoring system based on Aider coding agent\footnote{\url{https://aider.chat}} and a dynamic user leaderboard~\footnote{\href{https://mera-evaluation.github.io/demo-swe-mera/leaderboard}{SWE-MERA leaderboard}}\footnote{The video screencast of the user's journey can be accessed through the \href{https://youtu.be/1NoJh-e0uTk?si=FMmAIBgotKQOIzn4}{link} provided}. 
\end{enumerate}

\section{Related Work}
SWE-bench introduced a semi-automatic pipeline for mining software engineering tasks from popular open-source Python repositories, resulting in a benchmark of 2,294 issues and corresponding pull requests. Although this enabled a large-scale evaluation, the dataset suffered from quality issues, including poorly specified tasks and weak test coverage, which compromised the reliability of model assessment. To improve data quality, SWE-bench Verified\footnote{\url{https://openai.com/index/introducing-swe-bench-verified/}} released a human-validated subset of 500 tasks from SWE-bench, but this approach has limited scalability. Further work, such as SWE-Bench+~\cite{aleithan2024swe}, revealed that a significant portion of the solutions in the original dataset could be ``cheated'' due to solution leakage in the issue or pull request descriptions, highlighting the risks of data contamination, as most issues predated significant LLM knowledge cutoffs. SWE-Bench+ addressed these issues by filtering for post-cutoff tasks and removing instances with leaked solutions, resulting in a more robust benchmark.

To expand diversity and generalizability, Multi-SWE-bench~\cite{zan2025multi} extended coverage to multiple programming languages. Complementary approaches, such as SWE-Gym~\cite{pan2024training} and SWE-smith~\cite{yang2025swe}, focused on automatic task generation and scalable synthetic data creation, respectively, to further increase the size and diversity of a benchmark.

While these repository-level benchmarks advance the field, they remain largely static or require substantial manual curation. In contrast, LiveCodeBench~\cite{jain2024livecodebench} pioneered a dynamic, frequently updated evaluation framework to address contamination. However, it primarily targets algorithmic problems and does not capture the repository-level complexity essential for a realistic software engineering assessment.

\section{Methodology}

The SWE-MERA task collection process systematically generates evaluation tasks based on real-world software engineering challenges. A comprehensive parsing of publicly available repositories was conducted to maximize coverage. For each selected repository state, tests were identified that are introduced in subsequent development but do not yet pass in the current version.

This framework enables objective assessment: after reverting the repository to the specified state and incorporating these future test cases, tests categorized as \texttt{PASS\_TO\_PASS} are expected to \textit{succeed}, while those labeled \texttt{FAIL\_TO\_PASS} are expected to \textit{fail}.

\begin{table*}[tbh!]
    \centering
    \resizebox{\linewidth}{!}{
    \begin{tabular}{lccc}
        \toprule
        \textbf{Step} & \textbf{Total Repositories} & \textbf{Total Issues} & \textbf{Time Estimation} \\
        \midrule
        GitHub \\
        \:\:\:\:all public
            & 
            255M 
            & 
            522M
            & 1 sec. \\
        \:\:\:\:\:\: Python
            & 
            21M 
            & 
            53M 
            & 1 sec. \\
        \:\:\:\:\:\:\:\: 10+start, 10+forks & 
        168K 
        & 
        5.7M 
        & 7 hours \\
        \:\:\:\:\:\:\:\:\:\: 1+ closed issue
        & 
         97K
        & 5.7M (4.2M$^\dagger$)
        & 7 hours \\
        Repository Selection \\
        \:\:\:\:\:Python, 10+start, 10+forks, repo updated at 2025
        & 110K 
        & 5.5M 
        & 7 hours \\
        PR-Issue Mapping Construction  \\
        \:\:\:\:\:1+ updated issue at [2025-01-01, 2025-06-01]
        & 
        25K
        & 
        339K
        & 3 days \\
        \:\:\:\:\:\:\:\:issue is closed, closed merge request
        & 
         10K
        & 
         98K
        & 3 days \\
        \:\:\:\:\:\:\:\:\:\: one-to-one mapping beetween issse and pr 
        & 8.4K
        & 55K
        & 2 min. \\
        Metadata Extraction and Filtering
        & 8.2K
        & 51K
        & 11 hours \\
        Patch Extraction and Validation
        & 6.7K
        & 30K
        & 12 hours \\
        Repository Build Validation
        & 1.6K
        & 9K
        & 3 hours 
        \\
        End-to-End Task Execution
        & 668 
        & 1.6K
        & 2 days \\
        LLM-based pipeline evaluation & 279 & 528 & 2 hours \\
        \bottomrule
    \end{tabular}
    }
    \caption{
    Summary of the task collection funnel for the period 2025-01-01 to 2025-06-01, calculated using the GitHub GraphQL API. For our experiments, Repository Build Validation was performed immediately before PR-Issue Mapping Construction to minimize total processing time; this table is provided for reference.\\
    $^\dagger$ Only closed issues.
    }
    \label{tab:pipeline}
\end{table*}

\subsection{Steps overview}

To ensure the transparency and reproducibility of the task generation process, we have designed a well-documented and accessible collection pipeline. 

This pipeline is executed on a monthly basis, enabling the systematic and continuous collection of tasks over time. By adhering to this schedule, we facilitate the regular updating and expansion of our dataset, ensuring that it remains current and reflective of ongoing developments in the software engineering domain.

In practice, we plan to execute the pipeline on a quarterly basis, with each run collecting tasks for the preceding three months (month by month).

The pipeline comprises the following steps:
\begin{enumerate}[nosep]
    \item \textbf{Repository Selection:} GitHub repositories are selected based on predefined criteria, including a minimum threshold of 10 stars and 10 forks, recent activity within the current year, Python as the primary programming language, and the presence of an open-source license.
    \item \textbf{PR-Issue Mapping Construction:} Mappings between issues and pull requests are constructed according to the following criteria:
\begin{itemize}
    \item Each pull request is associated with exactly one issue (either linked directly or via comments).
    \item Each issue is associated with exactly one pull request.
    \item The pull request is merged.
    \item The associated issue is closed.
    \item The pull request merge date is later than the first day of the previous month.
\end{itemize}

    
    \item \textbf{Metadata Extraction and Filtering:} For each selected issue and its corresponding pull request, metadata (including title, text, and comments) is downloaded and parsed. The issue-PR pairs are then filtered out if the combined length of the issue title and the issue body is less than 25 characters.
    
    \item \textbf{Patch Extraction and Validation:} For each pull request, the corresponding \texttt{git diff} is generated and validated. Only examples that modify both source code and test files are retained. Additionally, only pull requests that modify fewer than 15 source files are considered.
    
    \item \textbf{Repository Build Validation:} For each task, we build an appropriate environment in a Docker container.  Validation is considered successful if \texttt{pytest} returns at least one passed test.
    
    \item \textbf{End-to-End Task Execution:} Each generated task is executed in a controlled environment within a Docker container to verify its reproducibility and correctness. A detailed description can be found in Appendix~\ref{app:build_validation}.
    \item \textbf{LLM-based Pipeline Evaluation:} \label{step:llm_eval}
    To assess the quality of each candidate task, we use the Qwen3-32B model \cite{yang2025qwen3} to evaluate the description, patch, and associated tests on four criteria: \texttt{task correctness}, \texttt{test correctness}, \texttt{test completeness}, and \texttt{complexity}. The model is prompted to return a structured JSON response with a score of 1 to 10, a confidence value (0.0–1.0), and a brief explanation for each criterion (see Appendix~\ref{app:llm_prompt} for the prompt used).

    We filter out tasks that fall into the \textit{bottom quartile} (lower 25\%) of the score distribution for any of the following dimensions:
    \begin{itemize}
        \item \texttt{task correctness}
        \item \texttt{test correctness}
        \item \texttt{test completeness}
    \end{itemize}
    The \texttt{complexity} score is not used for filtering, as we explicitly aim to retain both easy and difficult tasks. This filtering step ensures that retained tasks are well-formed, solvable, and adequately tested.
    
\end{enumerate}

The detailed results of each pipeline step are summarized in Tab.~\ref{tab:pipeline}. The sample collected task can be found in Appendix~\ref{app:task_example}.






\subsection{Availability}

The entire pipeline is implemented as a Python package\footnote{\url{https://pypi.org/project/repositorytest}; source code for the package can be found \href{https://github.com/MERA-Evaluation/repotest}{here}.}  and can be executed for any GitHub repository, facilitating reproducibility and extensibility for future research.

All tasks are typically executed within Docker containers using a standardized base image\footnote{\href{https://hub.docker.com/layers/library/python/3.11/images/sha256-6743feec06fa92c015a54dcc0f05cb06bef866dfb3e440b0485b4b9b27b6e28a}{https://hub.docker.com/layers/library/python/3.11}}.


However, the same execution process can be replicated in a Conda environment without modification to the underlying code.

\section{Evaluation}

\begin{figure*}[htb!]
    \centering
    \includegraphics[width=\textwidth]{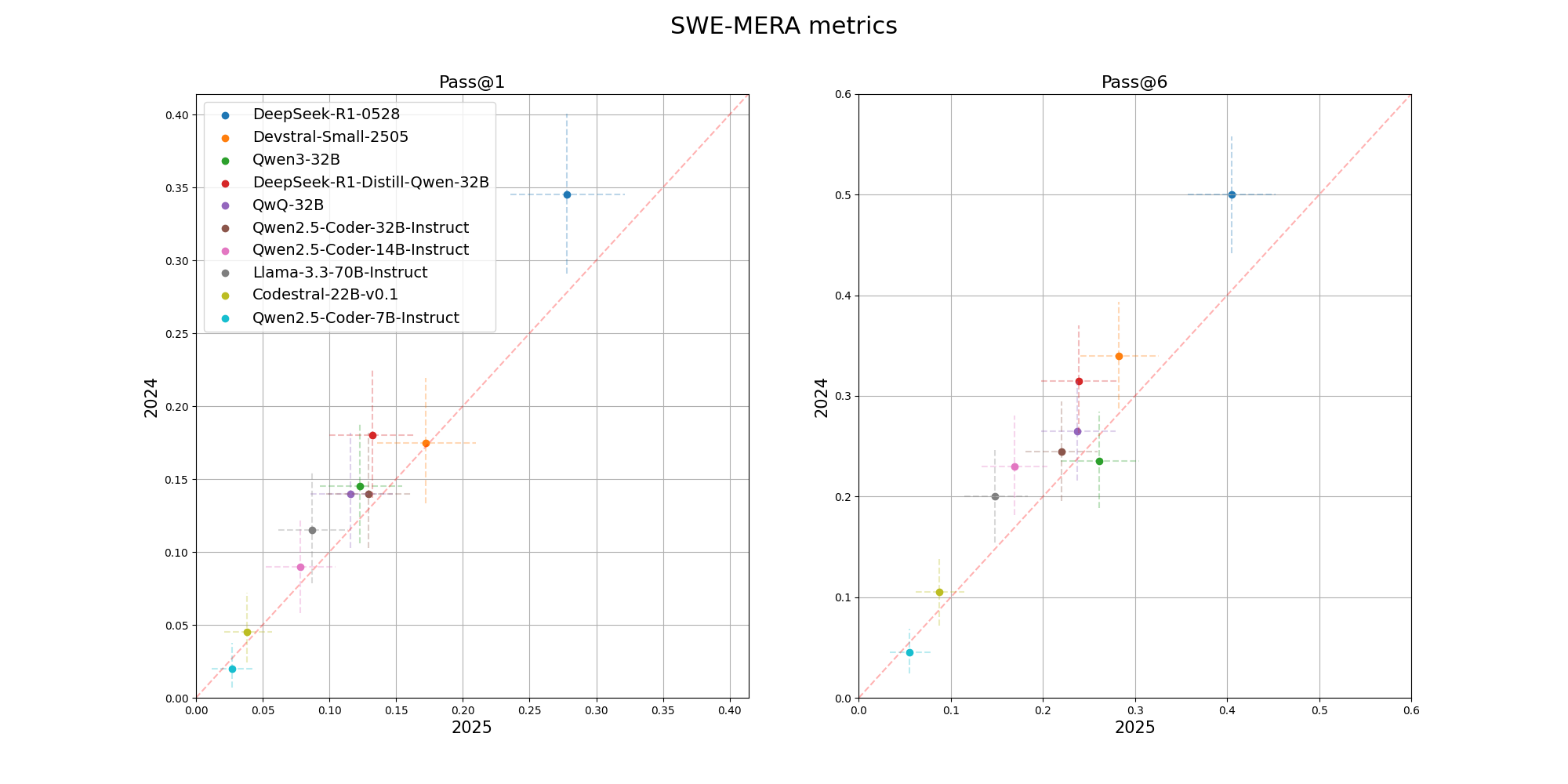}
    \caption{Comparison of model pass@1/pass@6 metrics between two years. Error bars represent confidence intervals, computed using the binomial distribution (5\% two-sided quantile).}
    \label{fig:pass1-pass6}
\end{figure*}

To apply LLMs in issue-solving scenarios, we employ a popular agentic framework Aider, which performs similarly to other state-of-the-art frameworks when applied to the same models. Aider gives six attempts (``tries'') to LLMs to fix a given issue, while every attempt allows up to four reflections to lint or test output. We slightly modify Aider~\footnote{\href{https://github.com/Aider-AI/aider/tree/4f4b10fd868680e0b87511d4bcf755f198089e8d}{github.com/Aider-AI/aider@4f4b10fd}}
and Aider-SWE-bench%
\footnote{\href{https://github.com/Aider-AI/aider-swe-bench/tree/6e98cd6c3b2cbcba12976d6ae1b07f847480cb74}{github.com/Aider-AI/aider-swe-bench@6e98cd6}} repositories due to backward compatibility issues. However, we aim to support several popular frameworks capable of solving issues in the near future.

Aider gives \textit{six} consequent independent attempts to LLMs to fix issues, but if an LLM succeeds in early attempts, it moves to the next issue. Due to the experimental design, we report two metrics: pass@1, indicating whether the first attempt was successful, and pass@6, indicating whether any of the six attempts succeeded.

To assess benchmark reliability, we selected several popular state-of-the-art LLMs for code, including Qwen, Devstral, DeepSeek-R1, and others (see the full list in the next section). We ran these models on 8 NVIDIA H100 80 GB with exceptions for DeepSeek-R1 and 7B models run on 16 and 4 GPUs, respectively. The evaluation for a single model took, on average, 3$\pm$1 hours for Aider runs and half an hour to test final patches. This experiment used 60-140M prompt tokens and 3-20M completion tokens, corresponding to 14-20K prompt and 1-4K completion tokens per request. 

\input{tables/results}

\subsection{Evaluated Models}

     Codestral-22B-v0.1\footnote{\url{https://mistral.ai/news/codestral}} is a model trained by Mistral AI on a diverse dataset of 80+ programming languages, including the most popular ones, such as Python, Java, C, C++, JavaScript, and Bash. 
    
     Qwen2.5-Coder-{7,14,32}B-Instruct models~\cite{hui2024qwen2} are the latest Qwen LLMs designed for code, available in multiple parameter sizes, affording flexibility between resource usage and performance. Qwen2.5 Coder models significantly improve in code generation, code reasoning, and code fixing, and have a more comprehensive foundation for real-world applications such as Code Agents.
    
     Llama-3.3-70B-Instruct\footnote{\url{https://www.llama.com/docs/model-cards-and-prompt-formats/llama3_3/}} multilingual large language model (LLM) is an instruction-tuned generative model. The Llama 3.3 instruction-tuned text-only model is optimized for multilingual dialogue use cases. Note that it is a general-purpose chat model, not specifically designed for code. 
    
    DeepSeek-R1-0528 by DeepSeek AI~\cite{deepseekai2025deepseekr1incentivizingreasoningcapability} incorporates computational enhancements and novel post-training optimizations to significantly improve reasoning, inference, and problem-solving capabilities. The updated model achieves state-of-the-art benchmark performance with reduced hallucination rates while advancing code generation and function calling. As open-source software, it democratizes access to advanced reasoning capabilities.
    
     DeepSeek-R1-Distill-Qwen-32B is a Qwen2.5 32B model~\cite{qwen2.5} distilled on the reasoning data generated by DeepSeek-R1 \cite{deepseekai2025deepseekr1incentivizingreasoningcapability}. The distilled models demonstrated exceptionally high performance on other benchmarks.
    
     QwQ-32B is a reasoning model developed by the Qwen Team~\cite{qwq32b}, which achieves competitive performance compared to state-of-the-art reasoning models.

     Qwen3-32B~\cite{yang2025qwen3} supports 119 languages and features a unique dual-mode architecture enabling efficient switching between complex reasoning (``thinking mode'') and dialogue (``non-thinking mode''). This architecture delivers significant performance improvements in reasoning, code generation, and creative dialogue, surpassing prior Qwen models. Qwen3 also demonstrates leadership in agent integration and multilingual task performance.
    
     Devstral-Small-2505\footnote{\url{https://mistral.ai/news/devstral}\label{fn:devstral}}
is an open-source agentic language model for software engineering, developed by Mistral AI and All Hands AI, excelling at codebase exploration, multifile editing, and software engineering tool usage. 

We provide more information on the baselines in Appendix~\ref{app:baselines}.

\section{Results} 
\label{sec:results}

Tab.~\ref{tab:results} presents the evaluation results for state-of-the-art LLMs on the 2025 subset of SWE-MERA using the Aider agent workflow. The results indicate that SWE-MERA accurately ranks the Qwen2.5-Coder models according to their size. In addition, Qwen3-32B \cite{yang2025qwen3} slightly outperforms QwQ-32B, which is consistent with the declared model specifications. In particular, Devstral-Small-2505 demonstrates superior performance as reported in its release notes\footref{fn:devstral}, despite its smaller size.

We have found an interesting feature while comparing the results for the evaluation of the baselines for 2024 and 2025. It seems that DeepSeek-R1, both the 0528 and Distill-Qwen-32B versions, perform better on 2024 tasks. The other investigated models do not show such behaviour. The results are visualized in Fig.~\ref{fig:pass1-pass6}. More detailed results are presented in Appendix~\ref{app:results}.

\section{Discussion}

\paragraph{Scaling} We observed that the GitHub API rate limits are sufficient to collect all relevant tasks from the past month using a single GitHub token in two days, which is surprisingly fast. 

Currently, we have collected approximately 700 tasks. If we do not restrict our benchmark to the last 6 months, we estimate the number of collected tasks to be 10,000; however, achieving this scale will require additional effort, particularly for End-to-End execution tasks.

\paragraph{Malicious Software} In our security assessment, we scanned 668 repositories for known virus signatures and discovered two repositories that, while suitable for SWE benchmarking, also exhibited the signatures:
\begin{itemize}
    \item \url{https://github.com/DataDog/guarddog} An open-source security scanner designed to detect vulnerabilities and malicious dependencies in software supply chains.
    \item \url{https://github.com/fkie-cad/socbed}  A framework for simulating and evaluating security operations center (SOC) environments, supporting research in cyber defense and attack scenarios.
\end{itemize}
This finding underscores the need to integrate basic virus signature checking into our system to ensure the integrity and safety of the collected repositories.




\begin{figure*}[!ht]
    \centering
    \includegraphics[width=0.9\textwidth]{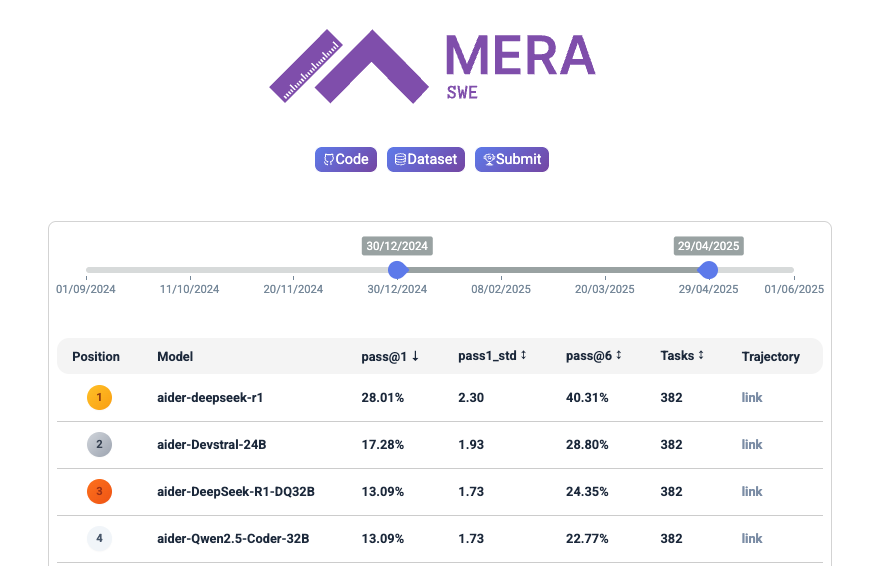}
    \caption{Screenshot of the SWE-MERA evaluation platform web interface.}
    \label{fig:submission_workflow}
\end{figure*}

\paragraph{Complexity} Our experiments demonstrate that the last step of our pipeline, namely the LLM-based evaluation, is crucial to maintain task quality. Automated collection methods may yield tasks that are either excessively complex due to insufficient information in the issue description or overly trivial when the solution is explicitly provided. The LLM-based assessment effectively filters such cases, ensuring that only tasks of appropriate complexity and relevance are retained.


\section{System Demonstration}

The SWE-MERA evaluation platform provides a reproducible and transparent environment to benchmark software engineering agents. The benchmark can be accessed at the \href{https://mera-evaluation.github.io/demo-swe-mera/}{link}. The web interface features an interactive slider, enabling users to visualize evaluation metrics across different dates and to inspect potential contamination events in the dataset, as shown in Fig.~\ref{fig:submission_workflow}.

\paragraph{Submission Workflow}

To participate in the evaluation and have your agent's results displayed on the leaderboard, one should follow these steps:

\begin{enumerate}[nosep]
    \item \textit{Dataset Acquisition:} Download the SWE-MERA dataset from the Hugging Face repository\footnote{\url{https://huggingface.co/datasets/MERA-evaluation/SWE-MERA}}.
    \item \textit{Agent Execution:} Run a software engineering agent on the provided dataset.
    \item \textit{Submission:} Submit the results by creating a pull request to the evaluation repository.
    \item \textit{Validation and Leaderboard Update:} Submissions are reviewed and, within two working days, valid results are integrated into the public leaderboard.
\end{enumerate}

A schematic overview of the submission process is provided in Fig.~\ref{fig:submission_workflow}.

\paragraph{Dataset and Evaluation Updates}

The SWE-MERA dataset is updated monthly and is available through the Hugging Face platform. Participants are encouraged to include links to their agent's execution trajectories; otherwise, the system will default to displaying data from the corresponding GitHub pull request.

\paragraph{Leaderboard and Data Visibility}

The dashboard is automatically updated to reflect new submissions. If a model does not have sufficient data to compute evaluation metrics for a selected time period, it will not be displayed on the leaderboard for that interval.

\paragraph{Interface Features}

\begin{itemize}[nosep]
    \item The web interface includes a slider for temporal navigation of metrics.
    \item Users can inspect detailed evaluation results and identify potential overfitting or contamination issues.
    \item The system supports transparent and reproducible evaluation, with all data and code accessible via public repositories.
\end{itemize}

\section{Conclusion}
SWE-MERA introduces a new approach to evaluating software engineering tasks, effectively addressing key limitations through dynamic data collection, automated quality validation, and ongoing updates to the dataset. This method helps mitigate concerns about data contamination while improving both task quality and the reliability of evaluations.

The benchmark demonstrates strong discriminative power across state-of-the-art models and establishes reliable performance baselines that are free from the contamination issues often found in traditional static benchmarks. With its extensible design and community-driven approach, SWE-MERA serves as a vital resource to advance AI research in software engineering.

The framework's adaptable design allows for expansion to programming languages such as Java, JavaScript, TypeScript, Go, and C++ by employing a standardized metadata approach. Future developments will focus on improving visualization capabilities, refining quality metrics, and integrating more closely with intelligent coding systems.


    
\section*{Limitations}

While the dynamic collection of coding problems in our benchmark framework presents distinct advantages, it also introduces several important limitations.

First, dynamically collected tasks, while allowing for scalability and novelty, may lack the nuanced complexity and creativity found in carefully curated or human-authored problems. Automatically constructed problems may inadvertently result in unnaturally phrased prompts, incomplete specifications, or tasks that are either trivial or excessively convoluted, which can compromise the validity and diversity of the benchmark.

Second, evaluating model performance on dynamically generated problems poses challenges for ground truth and grading quality. Automated reference solutions and test cases may not exhaustively capture all correct or optimal solutions, especially for open-ended or ambiguous problems. As a result, our metrics may underestimate model capabilities on creative or alternative approaches, and automated correctness checks may yield false negatives.

Third, ensuring the quality and fairness of dynamically collected problems is inherently difficult. It is possible for the generation process to introduce biases, such as overrepresenting certain programming paradigms, languages, or styles while underrepresenting others. This may affect the generalizability of evaluation results and obscure weaknesses of LLMs in underrepresented domains.

Fourth, although dynamic generation reduces risks of memorization and contamination from training data, it does not wholly eliminate them. For models trained on vast internet datasets, generated problems may still resemble well-known canonical challenges or textbook exercises, and thus performance may reflect prior exposure rather than true generalization abilities.

Fifth, the infrastructure for dynamic problem generation and grading brings additional technical complexity and potential instability. Failures in problem construction, test case generation, or sandboxed code execution can introduce noise into evaluation results and limit the reproducibility of benchmarking runs.

Finally, our current benchmark focuses primarily on programming correctness. Other crucial aspects of software engineering — such as code readability, maintainability, efficiency, security, and teamwork — are not evaluated in this framework and remain open challenges for future work.

\section*{Ethical Statement}

The introduction of a dynamically collected benchmark for evaluating LLM coding abilities raises several ethical considerations.

First, all prompts, solutions, and test cases produced by the dynamic generation system have been constructed to avoid the unintentional inclusion of proprietary, copyrighted, or sensitive information. The generation process is based solely on open-source templates, algorithmic patterns, and public domain resources, minimizing the risk of intellectual property infringement.

Second, although dynamically generated problems reduce the risks of data contamination and memorization in models, they do not fully mitigate the potential for LLMs to generate unsafe, insecure, or malicious code. We urge users to apply the benchmark ethically and to avoid using it—or the resulting models—for uses that may cause harm or violate responsible AI guidelines.

Third, by making dynamic generation tools and evaluation infrastructure publicly available, we strive to foster transparency, reproducibility, and equitable access to research resources. However, we recognize the potential for technology misuse, including the generation of synthetic coding tests for automated cheating on educational platforms or biasing LLM performance reviews for commercial interests. We recommend responsible stewardship, encourage open discussion of these risks, and welcome feedback from the broader community.

Fourth, while programmatically generated problems have clear scalability and adaptability benefits, there are potential risks of unintended bias in the selection or phrasing of tasks, which could disadvantage certain groups or languages. We commit to ongoing evaluation and refinement of the benchmark to ensure fairness, inclusivity, and diversity in the problems presented.

Lastly, we note that the widespread adoption of automated coding benchmarks has implications for education, employment, and the wider software ecosystem. Benchmarks should augment, rather than replace, comprehensive, human-centric evaluation of programming skills and ethical development practices.

\paragraph{AI-assistants Help} We improve and proofread the text of this article using Writefull assistant integrated in Overleaf (Writefull's/Open AI GPT models) and GPT-4o\footnote{\url{https://chatgpt.com}}, Grammarly\footnote{\href{https://app.grammarly.com/}{https://app.grammarly.com/}} to correct grammatical, spelling, and style errors and paraphrase sentences. We underline that these tools are used strictly to enhance the quality of English writing, in full compliance with the ACL policies on responsible use of AI writing assistance.
Nevertheless, some segments of our publication can be potentially detected as AI-generated, AI-edited, or human-AI-generated.

\bibliography{custom}

\appendix

\section{Repository Build Validation Procedure}
\label{app:build_validation}

The detailed validation process consists of the following steps:

\begin{enumerate}
    \item \textbf{Environment setup:}
    {\footnotesize
    \begin{lstlisting}[language=bash, basicstyle=\ttfamily\scriptsize]
pip install . && \
pip install pytest pytest-json-report
    \end{lstlisting}
    }
    \item \textbf{Test execution:}
    \begin{lstlisting}[language=bash, basicstyle=\ttfamily\scriptsize]
pytest --json-report \
 --json-report-file=report_pytest.json
    \end{lstlisting}

    \item \textbf{Success criteria:} Validation succeeds if:
    \begin{itemize}
        \item The command completes without errors.
        \item The JSON report shows more than 0 passed tests.
    \end{itemize}
    
    \item \textbf{Artifact preservation:} On success:
    \begin{itemize}
        \item The Dockerfile is saved for future image rebuilding.
        \item Docker cache is optimized for fast container recreation.
    \end{itemize}
\end{enumerate}

    


\section{Baselines}
\label{app:baselines}
More details of the baselines used are provided in Tab.~\ref{tab:baselines}.

\begin{table*}[ht!]
    \setlength{\tabcolsep}{3pt}
    \centering
    \resizebox{\linewidth}{!}{
    \begin{tabular}{m{6cm}>{\raggedleft\arraybackslash}m{3cm}>{\raggedleft\arraybackslash}m{3.5cm}>{\raggedleft\arraybackslash}m{2cm}}
    \hline
    \textbf{Model} &\textbf{Size} &\textbf{Release Date}& \textbf{Designed for code}\\
    \hline
    Codestral-22B-v0.1 & 22B & May 29, 2024 & yes\\
    Qwen2.5-Coder-\{7,14,32\}B-Instruct & 7B, 14B, 32B & November 12, 2024 & yes\\
    Llama-3.3-70B-Instruct & 70B & December 6, 2024 & no\\
    DeepSeek-R1-Distill-Qwen-32B & 32B & January 20, 2025 & no\\
    QwQ-32B & 32B & March 5, 2025 & no\\
    Qwen3-32B & 32B & April 28, 2025 & no\\
    Devstral-Small-2505 & 24B & May 21, 2025 & yes\\
    DeepSeek-R1 & 671B (37B active) & May, 28, 2025 & no \\
    \hline
    \end{tabular}
    }
    \caption{Evaluated models specification.}
    \label{tab:baselines}
\end{table*}

\section{Additional Results}
\label{app:results}
\begin{figure}[htb!]
    \centering
    \includegraphics[width=\linewidth]{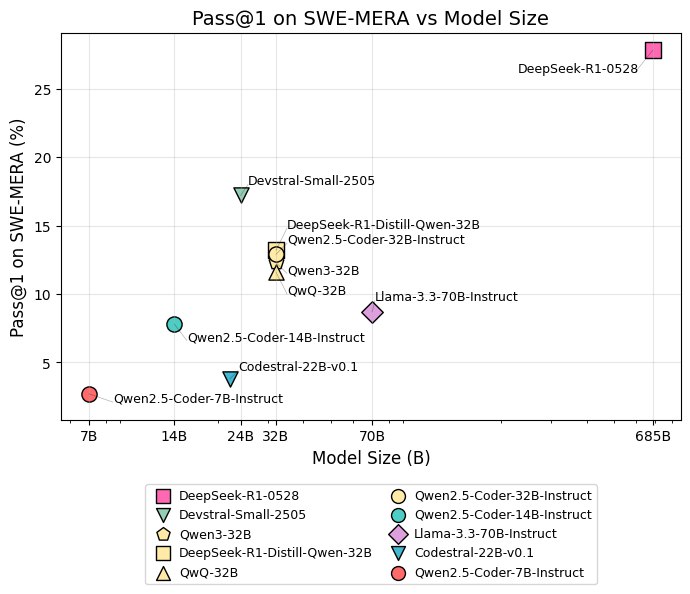}
    \caption{Pass@1 results vs model size for all evaluated models.}
    \label{fig:pass1-vs-model-size}
\end{figure}

\begin{figure}[htb!]
    \centering
    \includegraphics[width=\linewidth]{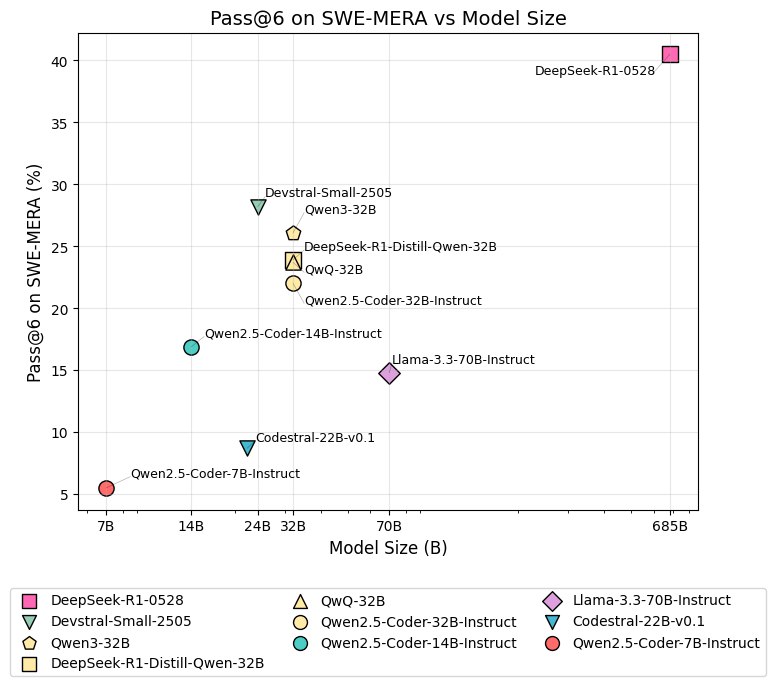}
    \caption{Pass@6 results vs model size for all evaluated models.}
    \label{fig:pass6-vs-model-size}
\end{figure}

Figs.~\ref{fig:pass1-vs-model-size}\&\ref{fig:pass6-vs-model-size} depict the results of the models compared to their respective sizes. The results here are the same as those in Tab.~\ref{tab:results}. 

Tab.~\ref{tab:results24} contains results of the evaluation on tasks dated 2024 year only. A year-over-year comparison (Tab.~\ref{tab:results} and~\ref{tab:results24}) shows that DeepSeek-R1 decreases from 50\% to 40.2\%, which is larger than all other models on average.
Devstral-Small-2505 from 34\% to 28.2\%, DeepSeek-R1-Distill-Qwen-32B from 31.5\% to 23.9\%, and Llama-3.3-70B-Instruct from 20\% to 14.8\%. Moreover, Qwen2.5-Coder-14B-Instruct achieves a 23\% pass@6 rate on 2025 data, which is similar to the results obtained by 32B models, whereas its pass@1 rate remains notably lower than that of the larger models. 

\begin{table*}[tbh!]
    \centering
    \resizebox{\linewidth}{!}{
    \begin{tabular}{lrrrrrr}
    \hline
    \textbf{Model} &\textbf{ pass@1 } &\textbf{ pass@6 }& \textbf{localize files} & \textbf{generate patch} & \textbf{regression tests} & \textbf{token limit hit}\\
    \hline
    DeepSeek-R1-0528 & 
    \textbf{34.5\%} & \textbf{50.0\%} & 90.9\% & 98.0\% & 49.7\% & 0.0\% \\
    Devstral-Small-2505 & 
    17.5\% & 34.0\% & 92.0\% & 99.0\% & 33.5\% & 0.5\% \\
    DeepSeek-R1-Distill-Qwen-32B & 
    18.0\% & 31.5\% & 89.4\% & 98.5\% & 32.7\% & 0.0\% \\
    QwQ-32B & 
    14.0\% & 26.5\% & 81.5\% & 96.5\% & 25.5\% & 0.5\% \\
    Qwen2.5-Coder-32B-Instruct & 
    14.0\% & 24.5\% & 91.5\% & 98.0\% & 24.6\% & 7.0\% \\
    Qwen3-32B & 
    14.5\% & 23.5\% & 94.5\% & 96.5\% & 25.1\% & 1.0\% \\
    Qwen2.5-Coder-14B-Instruct & 
    9.0\% & 23.0\% & 85.0\% & 95.0\% & 22.5\% & 3.0\% \\
    Llama-3.3-70B-Instruct & 
    11.5\% & 20.0\% & 84.4\% & 79.9\% & 19.6\% & 0.0\% \\
    Codestral-22B-v0.1 & 
    4.5\% & 10.5\% & 81.5\% & 84.5\% & 10.5\% & 3.5\% \\
    Qwen2.5-Coder-7B-Instruct & 
    2.0\% & 4.5\% & 68.8\% & 57.8\% & 4.0\% & 4.5\% \\
    \hline
    \end{tabular}
    }
    \caption{Evaluation results of models on SWE-MERA 2024. `localize files` is the percentage of attempts correctly identifying files to fix; `generate patch`, those producing valid patches; `regression tests`, those where patches pass original repository tests; and `token limit hit`, those exceeding the 32k token context limit.}
    \label{tab:results24}
\end{table*}

\begin{figure*}[htb!]
    \centering
    \includegraphics[width=\textwidth]{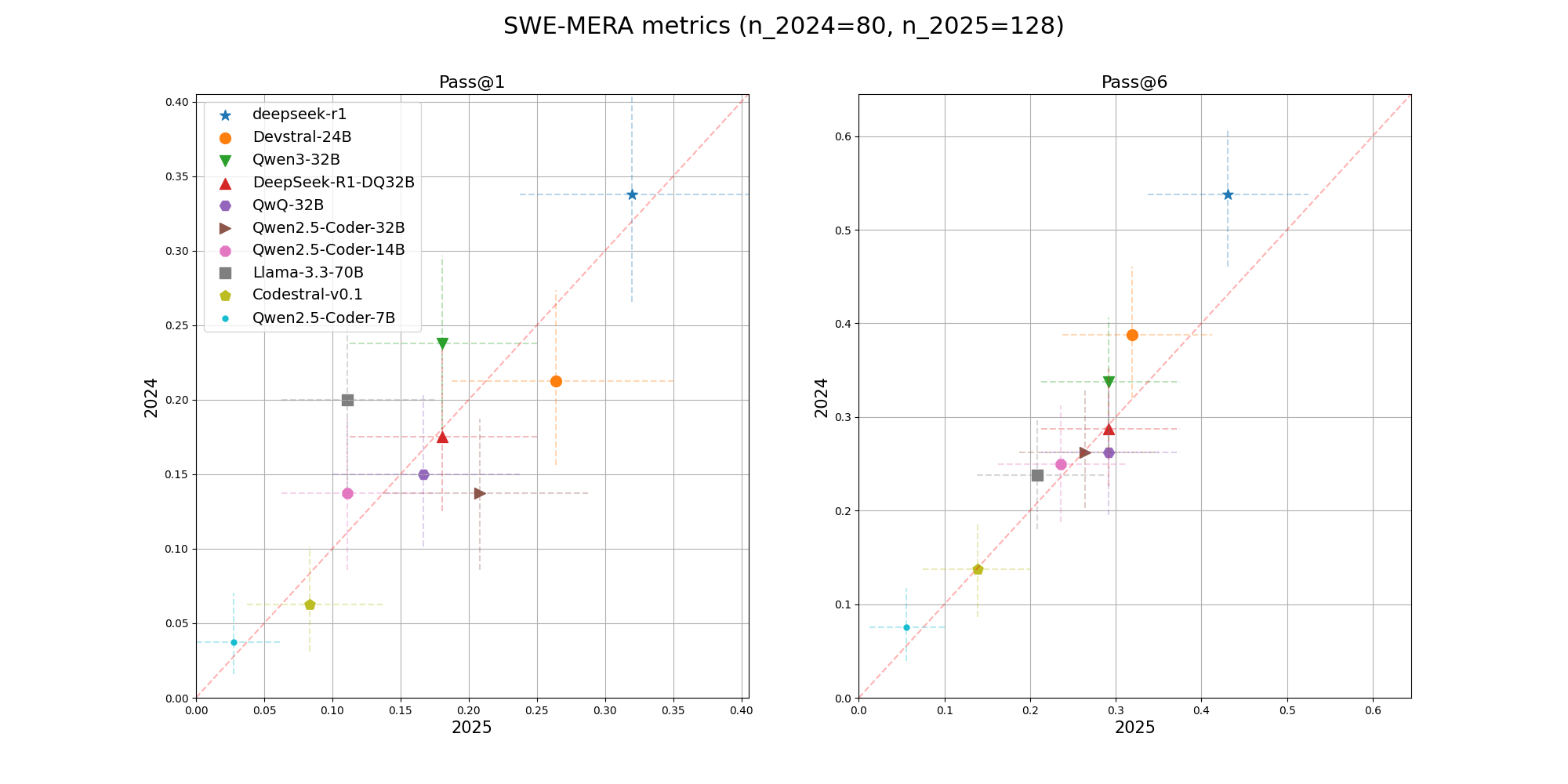}
    \caption{Comparison of model pass@1 and pass@6 metrics between two years. Error bars indicate confidence intervals, computed using the binomial distribution with a 5\% two-sided quantile. To address the variability in task sets across years, we include only those tasks for which $\min(\text{task\_correctness}, \text{test\_correctness}, \text{test\_completeness}) \geq 9$.}
    \label{fig:pass1-pass6-var}
\end{figure*}

Fig.~\ref{fig:pass1-pass6-var} represents the additional evaluation of the baselines in a more strict setup, where we take only the top decile (top 10\%) of all the tasks. Here, the year-to-year differences in behavior of the models are more subtle, namely, only DeepSeek-R1 shows a statistically valid discrepancy measured in pass@6.

\clearpage

\onecolumn
\section{Prompt for LLM-based Task Evaluation}
\label{app:llm_prompt}

We used the following prompt to evaluate the quality of candidate tasks using the Qwen3-32B model:

\begin{verbatim}
Conduct a comprehensive evaluation of the programming task solution
based on four criteria.

TASK:
{problem_statement}

SOLUTION:
{patch}

TESTS:
{test_patch}

Evaluate based on the following criteria:

1. TASK CORRECTNESS: Does the solution (patch) correctly solve the described problem?
2. TEST CORRECTNESS: Do the tests cover the problem from the task description?
3. COMPLEXITY: What is the complexity of solving this task?
4. TEST COMPLETENESS: Do the tests cover corner cases from the problem description?

Respond in JSON format:
{
    "task_correctness": {
        "score": <a score from 1 to 10>,
        "confidence": <a confidence score from 0.0 to 1.0>,
        "reasoning": "<explanation>"
    },
    "test_correctness": {
        "score": <a score from 1 to 10>,
        "confidence": <a confidence score from 0.0 to 1.0>,
        "reasoning": "<explanation>"
    },
    "complexity": {
        "score": <a score from 1 to 10>,
        "confidence": <a confidence score from 0.0 to 1.0>,
        "reasoning": "<explanation>"
    },
    "test_completeness": {
        "score": <a score from 1 to 10>,
        "confidence": <a confidence score from 0.0 to 1.0>,
        "reasoning": "<explanation>"
    }
}
\end{verbatim}

\section{SWE-MERA Task Example}
\label{app:task_example}
\input{data_samples/task_example}
\twocolumn


\end{document}

%% file: tables/results.tex
\begin{table*}
    \centering
\resizebox{\linewidth}{!}{
    \begin{tabular}{lrrrrrr}
    \hline
    \textbf{Model} &\textbf{ pass@1} &\textbf{ pass@6 }& \textbf{localize files} & \textbf{generate patch} & \textbf{regression tests} & \textbf{token limit hit}\\
    \hline
    DeepSeek-R1-0528 & 
    \textbf{27.8\%} & \textbf{40.2\%} & 89.6\% & 98.1\% & 40.6\% & 0.2\% \\
    Devstral-Small-2505 & 
    17.2\% & 28.2\% & 89.1\% & 98.7\% & 28.5\% & 0.4\% \\
    Qwen3-32B & 
    12.3\% & 26.1\% & 91.8\% & 98.9\% & 26.1\% & 1.1\% \\
    DeepSeek-R1-Distill-Qwen-32B & 
    13.2\% & 23.9\% & 87.8\% & 98.7\% & 23.7\% & 0.4\% \\
    QwQ-32B & 
    11.6\% & 23.7\% & 79.6\% & 96.6\% & 22.9\% & 0.6\% \\
    Qwen2.5-Coder-32B-Instruct & 
    12.9\% & 22.0\% & 86.3\% & 96.3\% & 22.0\% & 4.6\% \\
    Qwen2.5-Coder-14B-Instruct & 
    7.8\% & 16.9\% & 86.0\% & 93.7\% & 16.8\% & 1.9\% \\
    Llama-3.3-70B-Instruct & 
    8.7\% & 14.8\% & 77.0\% & 70.9\% & 14.8\% & 0.0\% \\
    Codestral-22B-v0.1 & 
    3.8\% & 8.7\% & 76.7\% & 83.4\% & 8.4\% & 2.9\% \\
    Qwen2.5-Coder-7B-Instruct & 
    2.7\% & 5.5\% & 58.2\% & 55.3\% & 4.8\% & 5.7\% \\
    \hline
    \end{tabular}
    }
    \caption{Evaluation results of models on SWE-MERA 2025. `localize files` is the percentage of attempts correctly identifying files to fix; `generate patch`, those producing valid patches; `regression tests`, those where patches pass original repository tests; and `token limit hit`, those exceeding the 32k token context limit.}
    \label{tab:results}
\end{table*}

%% file: data_samples/task_example.tex
\noindent\textbf{Problem Statement} \\
\rule{\columnwidth}{0.4pt}

\begin{verbatim}
Performance threshold goes to -inf when it should be zero.

In attempting to create a performance test where zero is the correct value, 
I created the following reference (since a value of zero results in no reference 
check being performed; see https://github.com/reframe-hpc/reframe/issues/2857):

    self.reference = {
        '*': {
            'Gflops': (None, None, None, 'Gflops'),
            'Exponent': (None, None, None, '10exp'),
            'Time': (None, None, None, 'seconds'),
            'failed_tests': (.1, -1.0, 0, 'tests'),
            'skipped_tests': (.1, -1.0, 0, 'tests')
    }}

I was thinking for the failed and skipped tests, this would create a lower bound 
of zero and upper bound of .1, allowing zero to pass, but integers larger than 
that would fail.
...
\end{verbatim}

\vspace{1em}

\noindent\textbf{Test Patch (Diff)} \\
\rule{\columnwidth}{0.4pt}

\begin{verbatim}
diff --git a/unittests/test_sanity_functions.py b/unittests/test_sanity_functions.py
index 7e6368938..0e9367027 100644
--- a/unittests/test_sanity_functions.py
+++ b/unittests/test_sanity_functions.py
@@ -473,6 +473,18 @@ def test_assert_reference():
                              r'\(l=-1\.2, u=-0\.9\)'):
         sn.evaluate(sn.assert_reference(-0.8, -1, -0.2, 0.1))

+    # Check that bounds are correctly calculated in case that lower bound
+    # reaches zero (see also GH issue #3430)
+    with pytest.raises(SanityError,
+                       match=r'1 is beyond reference value 0\.1 '
+                             r'\(l=0\.0, u=0\.1\)'):
+        assert sn.assert_reference(1, 0.1, -1.0, 0)

+    with pytest.raises(SanityError,
+                       match=r'-1 is beyond reference value -0\.1 '
+                             r'\(l=-0\.1, u=-0\.0\)'):
+        assert sn.assert_reference(-1, -0.1, 0, 1.0)

     # Check invalid thresholds
     with pytest.raises(SanityError,
                        match=r'invalid high threshold value: -0\.1'):
\end{verbatim}

\vspace{1em}

\noindent\textbf{Gold Patch (Diff)} \\
\rule{\columnwidth}{0.4pt}

\begin{verbatim}
diff --git a/reframe/utility/sanity.py b/reframe/utility/sanity.py
index a4f57a301..1228586ff 100644
--- a/reframe/utility/sanity.py
+++ b/reframe/utility/sanity.py
@@ -8,6 +8,7 @@
 import contextlib
 import glob as pyglob
 import itertools
+import math
 import os
 import re
 import sys
@@ -576,8 +577,14 @@ def calc_bound(thres):

         return ref*(1 + thres)

-    lower = calc_bound(lower_thres) or float('-inf')
-    upper = calc_bound(upper_thres) or float('inf')
+    lower = calc_bound(lower_thres)
+    if lower is None:
+        lower = -math.inf

+    upper = calc_bound(upper_thres)
+    if upper is None:
+        upper = math.inf
...
\end{verbatim}